# Beating the Gilbert-Varshamov Bound for Online Channels

Ishay Haviv*    Michael Langberg†

**Abstract**

In the online channel coding model, a sender wishes to communicate a message to a receiver by transmitting a codeword $x = (x_1, \ldots, x_n) \in \{0,1\}^n$ bit by bit via a channel limited to at most $pn$ corruptions. The channel is online in the sense that at the $i$th step the channel decides whether to flip the $i$th bit or not and its decision is based only on the bits transmitted so far, i.e., $(x_1, \ldots, x_i)$. This is in contrast to the classical adversarial channel in which the corruption is chosen by a channel that has full knowledge on the sent codeword $x$. The best known lower bound on the capacity of both the online channel and the classical adversarial channel is the well-known Gilbert-Varshamov bound. In this paper we prove a lower bound on the capacity of the online channel which beats the Gilbert-Varshamov bound for any positive $p$ such that $H(2p) < \frac{1}{2}$ (where $H$ is the binary entropy function). To do so, we prove that for any such $p$, a code chosen at random combined with the nearest neighbor decoder achieves with high probability a rate strictly higher than the Gilbert-Varshamov bound (for the online channel).

## 1 Introduction

The classical scenario in coding theory is that of a sender Alice who wants to transmit a message $u$ to a receiver Bob via a binary communication channel. To do so, Alice encodes her message $u$ into a codeword $x = (x_1, \ldots, x_n) \in \{0,1\}^n$ and sends it to Bob, who is expected to recover the message $u$. However, the channel is allowed to corrupt (possibly probabilistically) at most a $p$-fraction of the codeword, i.e., to flip at most $pn$ bits in $x$, for some $p \in [0,1]$. The goal is to find a coding scheme by which Alice can send as many distinct messages as possible while ensuring correct decoding by Bob with high probability (over the encoding, decoding and the channel). Roughly speaking, we say that a code achieves rate $R$ if $2^{Rn}$ distinct messages can be sent using codewords of length $n$. Viewing the channel as a malicious *jammer*, it is important to specify what information the channel has while deciding on which bits to flip. Such a specification defines the model of communication and strongly affects the obtainable rate of communication.

In one extreme, there is the *classical adversarial model* in which the channel has full knowledge on the entire transmitted codeword $x$. Given $x$ and the coding scheme of Alice and Bob, the channel chooses an error for $x$. Calculating the maximum achievable rate for such a channel is a fundamental open problem in coding theory. The best known lower bound on the rate is due to Gilbert [8] and Varshmov [20] and equals $1 - H(2p)$, where $H$ stands for the binary entropy function. Namely, Gilbert and Varshamov show that there exists a subset of $\{0,1\}^n$ of size roughly $2^{(1-H(2p))n}$ in which every two distinct vectors have Hamming distance at least $2pn + 1$. This implies that if we take the vectors in this set as codewords then a nearest neighbor decoder always recovers the correct sent codeword. On the other hand, the best known upper bound is due to McEliece et al. [15] and is strictly higher than the Gilbert-Varshamov bound for any $p \in (0, \frac{1}{4})$.

In the second extreme, there are channel models in which the error imposed on the codeword $x$ is completely independent of $x$. An example of such a channel is the well-known *binary symmetric channel* studied

*The Blavatnik School of Computer Science, Tel Aviv University, Tel Aviv 69978, Israel. Supported by the Adams Fellowship Program of the Israel Academy of Sciences and Humanities. Work done in part while at the Open University of Israel.

†The Computer Science Division, Open University of Israel, Raanana 43107, Israel. Email: mikel@openu.ac.il. Work supported in part by ISF grant 480/08 and by the Open University of Israel's research fund (grant no. 46114).

(among other channels) by Shannon [19]. In this channel every transmitted bit is flipped independently with probability $p$, no matter what the sent codeword is. As opposed to the classical adversarial model, the picture here is completely clear, since Shannon proved that $1 - H(p)$ is a tight lower and upper bound on the maximum achievable rate.

In this work we continue the line of research in [12, 6, 7] which study the *online channel model* — a channel model whose strength lies somewhere between the above two extremes. In the online channel model, Alice sends a codeword $x$ bit by bit over a binary communication channel. For each $1 \leq i \leq n$ the channel decides whether to flip the $i$th bit or not immediately after $x_i$ arrives. This means that the channel's decision depends only on $(x_1, \ldots, x_i)$. As in the adversarial model, the channel is limited to corrupt at most $pn$ of the bits. Roughly speaking, the online channel is *stronger* than the binary symmetric channel, as an online channel can mimic the random behavior of a binary symmetric channel. On the other hand, the online channel is *weaker* than the classical adversarial channel, as an online channel is limited to make its decisions in a causal manner. The main theme of this work is to better understand the strength of the online channel model — in particular, does the maximum achievable rate when communicating over online channels resemble that of the classical adversarial channel, that of the binary symmetric channel, or maybe neither?

Studying online adversarial channels is naturally motivated by practical settings in which the sent message is not known to the channel which simultaneously learns it. For example, the online channel model simulates a transmission of a codeword $x$ via $n$ uses of a channel over time, where at time $i$ the $i$th bit of $x$ is transmitted. At each step the channel decides whether to flip $x_i$ whereas the receiver waits until the end of the transmission before decoding. As in the classical adversarial channel model, the channel is limited to at most $pn$ corruptions, what is usually interpreted as limited processing power or transmit energy. From a theoretical point of view, understanding the online channel model and comparing it to the classical adversarial channel model might shed some light on the capacity of the classical adversarial channel, a long-standing open problem in coding theory.

## 1.1 Related Work

Let $C_{\text{online}}(p)$ denote the *capacity of the online channel*, defined as the maximum achievable rate when communicating over an online channel allowed to corrupt at most a $p$-fraction of the transmitted codeword. We give a rigorous definition of the capacity $C_{\text{online}}(p)$ in Section 2. The known bounds on the capacities of the classical adversarial channel and the binary symmetric channel immediately imply some bounds on the capacity of the online channel. It is clear that any coding scheme that works for the classical adversarial channel works also for the online channel, and hence $C_{\text{online}}(p) \geq 1 - H(2p)$. On the other hand, the online channel can flip every bit independently with probability $p$ (up to $pn$ of them) ignoring the transmitted codeword $x$. It is not hard to verify that this implies that Shannon's upper bound (for the binary symmetric channel) holds for the online channel model as well, that is, $C_{\text{online}}(p) \leq 1 - H(p)$. Recently, this upper bound was improved in [12] for any $p \geq 0.15642$. More precisely, it was shown in [12] that for any $p \geq \frac{1}{4}$ no communication with positive rate is possible via the online channel and that for $p < \frac{1}{4}$, $C_{\text{online}}(p) \leq 1 - 4p$. This implies that the online channel model is strictly stronger than the binary symmetric channel, in the sense that there exist values of $p$ (e.g., $p = \frac{1}{4}$) for which no communication is possible over the online channel whereas a positive rate is possible for the binary symmetric channel. In [12] no non-trivial lower bounds on $C_{\text{online}}(p)$ were presented. The state of the art on the online channel model is given below (see Figure 1).

**Theorem 1.1** ([12]). *For any* $p \in [0, \frac{1}{2}]$, *it holds that* $1 - H(2p) \leq C_{online}(p) \leq \min\left(1 - H(p), (1 - 4p)^+\right)$, *where* $(1 - 4p)^+$ *is defined to be* $1 - 4p$ *for* $p < \frac{1}{4}$ *and* $0$ *otherwise.*

The problem of coding against online channels over large alphabets was studied in [6], where a full characterization of the capacity is presented. Namely, it is shown in [6] that when communicating over large alphabets, the online channel is no weaker than the classical adversarial channel and has capacity $1 - 2p$ for $p < \frac{1}{2}$ and 0 otherwise. The proofs of the tight upper and lower bounds in [6] use the geometry that fields of large size enjoy, and it is not clear if these ideas can be extended to the binary case considered in our work.

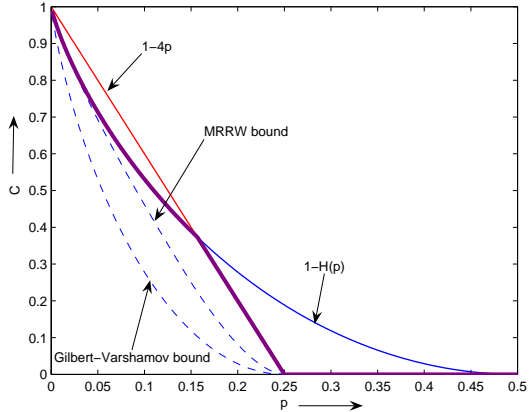

Figure 1: The bounds on the capacities of the classical adversarial channel and the online channel. The bold line (in purple) is the upper bound on the capacity of the online channel from [12].

To the best of our knowledge, other than the works mentioned above, communication in the presence of an online channel has not been explicitly addressed in the literature. Nevertheless, we note that the model of online channels, being a natural one, has been "on the table" for several decades and the analysis of the online channel model appears as an open question in the book of Csiszár and Korner [4] in the section addressing Arbitrarily Varying Channels (AVC) [2]. (The AVC model is a broad framework for modeling channels, which encapsulates our online model. For a nice survey on AVCs see [13].) In addition, various variants of online channels have been addressed in the past, for instance [2, 11, 17, 18, 16, 9] – however the models considered therein differ significantly from ours.

## 1.2 Our Result

The Gilbert-Varshamov rate of $1 - H(2p)$ is the state of the art when communicating over classical adversarial channels. The question whether one can improve upon this rate when communicating over online channels is an intriguing question. An affirmative answer would not only make progress in our understanding of the online channel model but also may hint on a possible separation between the online and classical adversarial channels. In our work we address this question and present a lower bound on the capacity of the online channel that beats the Gilbert-Varshamov bound. More precisely, we prove that for any small enough $p$, the Gilbert-Varshamov lower bound is not tight for the online channel. This means that for any such $p$, there exists a coding scheme for the online channel with rate strictly higher than $1 - H(2p)$. This is the first lower bound for the online channel which is not known to hold for the classical adversarial model. Our result is stated below.

**Theorem 1.2.** *For any $p$ such that $H(2p) \in (0, \frac{1}{2})$ there exists a $\delta_p > 0$ such that*

$$C_{online}(p) \geq 1 - H(2p) + \delta_p.$$

Note that $H(2p) \in (0, \frac{1}{2})$ for any $p \in (0, \frac{1}{2} \cdot H^{-1}(\frac{1}{2})) \approx (0, 0.055)$. We also note that our result holds with respect to the *average* error criteria (see Section 2.2 for a discussion on the error type). Finally, we remark that in order to prove Theorem 1.2 we show a lower bound on a much stronger channel model, which we refer to as the *two-step model* (defined below).

## 1.3 Techniques and Proof Overview

Our goal in this paper is to show the existence of an encoder and a decoder for the online channel by which Alice and Bob achieve some rate $R$ strictly higher than $1 - H(2p)$, which is the rate achieved by the Gilbert-Varshamov bound. Instead of dealing directly with the online channel model we consider a stronger channel model, the *two-step model*, defined as follows. Denote $\alpha = R - \varepsilon$ for some small $\varepsilon > 0$. In the *first step* Alice sends the first $\alpha n$ bits of her encoded message and the channel (after viewing this transmitted information) decides which bits to flip out of these $\alpha n$ bits. In the *second step* Alice sends the rest of the codeword and the channel (now with full knowledge on the sent codeword) decides which bits to flip out of the remaining transmission. The number of bits corrupted in the two steps together is limited to be at most $pn$. Notice that this model is stronger than the online channel model in the sense that any code allowing communication over the two-step model will also allow communication over our model of online channels. Indeed, any adversarial strategy of the online channel model implies a valid strategy for the two-step model achieving the exact same parameters. Therefore, in order to prove our lower bound on the capacity in Theorem 1.2 it suffices to consider the two-step model.

We turn to describe our construction of codes that allow communication over the two-step model with rate $R$ greater than $1 - H(2p)$. We first note that no linear code will suffice. Roughly speaking, this follows from the fact that each codeword $x$ in a linear code has exactly the same "neighborhood structure". Thus, when a linear code is used, the problems of communicating over channels with limited information regarding the codeword $x$ and those with full information are equivalent.[1] We thus turn to study codes which are not linear. A natural candidate is a code in which the codewords are chosen completely at random and the decoder is the *nearest neighbor* decoder. More precisely, we pick a code $C : [2^{Rn}] \to \{0,1\}^n$ such that for every $u \in [2^{Rn}]$ the codeword $C(u)$ is independently and uniformly chosen from $\{0,1\}^n$. Given such a code, Bob outputs a message $u' \in [2^{Rn}]$ that minimizes the Hamming distance between $C(u')$ and the received corrupted vector.

In order to prove our theorem, we show that the decoding succeeds with high probability no matter how the adversarial online channel behaves. The intuitive idea is the following. In the first step Alice sends a prefix $m \in \{0,1\}^{\alpha n}$ of a codeword where $\alpha = R - \varepsilon$. Since the code $C$ was constructed randomly, for a typical prefix $m$ there are exponentially many (about $2^{\varepsilon n}$) codewords in $C$ that share $m$ as a prefix. This means that the channel is not able to recognize the sent codeword at this point, and therefore it has no good way to decide which bits from $m$ to flip. Roughly speaking, we show that no matter which bits the adversary decides to flip in this first step, for *most* of the codewords that share $m$ as a prefix the error imposed by the adversary is in a *wrong* direction and thus will not enable the adversary to cause a decoding error (after the additional corruption of the second step). In fact, as our analysis shows, for our codes $C$ the best strategy for the adversary is actually to save its flipping power and to corrupt only in the second step of communication. This implies that in our setting the two-step channel will concentrate all its error on the second portion of the codeword! Comparing this state of affairs to the classical channel model in which the error is spread out over the entire codeword sheds light on the reason we are able to improve upon the Gilbert-Varshamov rate of $1 - H(2p)$. Very loosely speaking, to prove our improved rate, we first show that a code $C$ constructed at random *is expected* to allow successful communication. However, as the events corresponding to correct decoding are not independent of each other, our proof for the existence of the desired code follows a rather delicate analysis.

Our analysis holds for the two-step model and thus suffices to prove Theorem 1.2. To improve upon the results of Theorem 1.2, it is natural to try to generalize our analysis to a channel model that includes more than two steps. At its extreme (the $n$-step model) we obtain our original online channel. Such a generalized analysis is left open in this work and seemingly cannot be addressed by the current proof techniques.

---

[1]In detail, for any linear code of (minimum) distance at most $2pn$ there exists an online channel that causes any decoder to err with probability at least $\frac{1}{2}$ for every sent message. To see this, assume that $x$ and $y$ are two codewords of distance at most $2pn$, and let $z$ be a vector of distance at most $pn$ from both $x$ and $y$. Now, consider a channel that maps any codeword $w$ to $w + (z - y)$ or to $w + (z - x)$ with probability $\frac{1}{2}$ each. Observe that this is an online channel that causes any decoder to err with probability at least $\frac{1}{2}$.

In the following Section 2 we set the notation and definitions used throughout our work. We then turn to prove Theorem 1.2 in Section 3.

## 2 Preliminaries

### 2.1 Notations and Standard Definitions

For $k \in \mathbb{N}$ we denote $[k] = \{i \in \mathbb{N} \mid 1 \le i \le k\}$. For a vector $x = (x_1, \ldots, x_n) \in \{0,1\}^n$ and a number $1 \le k \le n$ we denote by $x|_{[k]}$ the projection of $x$ on its first $k$ entries, i.e., $x|_{[k]} = (x_1, \ldots, x_k)$. The *Hamming weight* of a binary vector is the number of its 1-entries, and the *Hamming distance* between $x \in \{0,1\}^n$ and $y \in \{0,1\}^n$, denoted by $\operatorname{dist}_H(x, y)$, is the Hamming weight of $x + y$, where the addition is modulo 2 and coordinate-wise.

For two functions $f, g : \mathbb{N} \to \mathbb{R}$, we say that $f$ and $g$ are *polynomially equivalent* and write $f \sim g$ if there are constants $c_1, c_2$ such that $n^{-c_1} \cdot f(n) \le g(n) \le n^{c_2} \cdot f(n)$ for all large enough $n \in \mathbb{N}$. Similarly, we write $f \lesssim g$ if there is a constant $c$ such that $f(n) \le n^c \cdot g(n)$ for all large enough $n \in \mathbb{N}$.

The *binary entropy function* $H : [0,1] \to [0,1]$ is defined by $H(0) = H(1) = 0$ and $H(p) = -p \log p - (1-p) \log(1-p)$ for $p \in (0,1)$, where the logarithms, here and everywhere in this paper, are of base 2. It is well-known and easy to verify that for any $c \in (0, \frac{1}{2}]$, $\binom{n}{cn} \sim 2^{H(c)n}$. We need the following two simple facts regarding $H$. Notice that the first fact implies the second (by setting the parameters of Fact 2.1 to be $x = 0$, $y = \frac{1}{2}$, and $\theta = 1 - 4p$).

**Fact 2.1.** *The entropy function $H$ is strictly concave, that is, for any $\theta \in (0,1)$ and $x, y \in [0,1]$ it holds that $\theta \cdot H(x) + (1-\theta) \cdot H(y) \le H(\theta \cdot x + (1-\theta) \cdot y)$, and equality holds if and only if $x = y$.*

**Fact 2.2.** *For any $p \in (0, \frac{1}{4})$, $4p < H(2p)$.*

We need the following version of the Chernoff-Hoeffding Bound [10, 14] (addressing random variables which are not necessarily indicator variables).

**Theorem 2.3** (Chernoff-Hoeffding). *Let $X_1, X_2, \ldots, X_N$ be independent and identically distributed random variables taking values in the unit interval $[0,1]$ with expectation at most $\mu$. Then,*

$$\Pr\left[\sum_{i=1}^N X_i \ge 2\mu N\right] \le e^{-\Theta(\mu N)}.$$

### 2.2 The Online Channel Model and the Two-step Model

For $R > 0$, an $(n, Rn)$-code $\mathbf{C}$ is a mapping $\mathbf{C} : [2^{Rn}] \to \{0,1\}^n$. The elements of the image of $\mathbf{C}$ are called *codewords*. Define $\alpha = R - \varepsilon$ for some $\varepsilon > 0$ and let $m \in \{0,1\}^{\alpha n}$ be some prefix. Here and throughout our work we ignore rounding issues and assume that $\alpha n$, $Rn$ and other such expressions are integers. We denote by $\mathbf{C}^m$ the set of all messages whose codewords have $m$ as a prefix, i.e., $\mathbf{C}^m = \{u \in [2^{Rn}] \mid \mathbf{C}(u)|_{[\alpha n]} = m\}$, and by $\overline{\mathbf{C}^m}$ the set of all messages whose codewords do not have $m$ as a prefix, i.e., $\overline{\mathbf{C}^m} = [2^{Rn}] \setminus \mathbf{C}^m$. A *random code* is a mapping $C : [2^{Rn}] \to \{0,1\}^n$ such that for every $u \in [2^{Rn}]$ the codeword $C(u)$ is independently and uniformly chosen from $\{0,1\}^n$. Notice that we use $\mathbf{C}$ to denote a fixed code and $C$ to denote a code which forms a random variable.

Consider a code $\mathbf{C}$. Throughout this work, we consider the *average error* success criteria while communicating over the online channel model. Namely, Alice's message $u$ is considered as uniformly distributed over $[2^{Rn}]$. Given the message $u$, Alice deterministically maps $u$ to the codeword $\mathbf{C}(u) = (x_1, \ldots, x_n) \in \{0,1\}^n$ and transmits it over the communication channel. For every $i \in [n]$ the decision of the channel whether to flip $x_i$ or not depends only on $(x_1, \ldots, x_i)$. In addition, the channel is limited to at most $pn$ corruptions. Bob's goal is to recover $u$ from his received vector.

The *probability of error* of **C** is defined as the average over all $u \in [2^{Rn}]$ of the probability of error for the message $u$, i.e., the probability that the message that Bob decodes differs from the message $u$ encoded by Alice. Here, the probability is taken over the random variables of the channel and of Bob. We say that the rate $R$ is *achievable* if for every $\varepsilon > 0$, $\delta > 0$ and every sufficiently large $n$ there exists an $(n, (R-\delta)n)$-code that allows communication with (average) probability of error at most $\varepsilon$. The supremum over $n$ of the achievable rates is called the *capacity of the online channel* and is denoted by $C_{\texttt{online}}(p)$. We note that the discussion in the introduction regarding the known bounds on the capacity of both the binary symmetric channel and the classical adversarial channel holds for average error (see e.g., [3]).

One may also consider a definition for capacity which takes into account the maximum error over messages $u$ and not the average error. In this maximum error (or *worst case*) setting, if the encoding function of Alice is considered to be deterministic, it is straightforward to verify that online channels have no advantage over the classical adversarial channel. This is no longer the case when one allows randomization in Alice's encoding process (referred to as *stochastic encoders*). As common in the study of Arbitrarily Varying Channels (e.g., [5]), there is an equivalence between the capacity when considering the models of (a) deterministic encoders and average error criteria and (b) stochastic encoders and maximum error success criteria. This equivalence holds also for the online channel model studied in this work.

As mentioned before, for our lower bound we consider a *two-step model* defined for a parameter $\alpha = R - \varepsilon$ where $\varepsilon > 0$ is some small constant. In the *first step*, Alice sends the first $\alpha n$ bits of the encoded message and the channel decides which bits to flip out of these $\alpha n$ bits. In the *second step*, Alice sends the rest of the codeword and the channel decides which bits to flip out of the remaining $(1-\alpha)n$ bits. The number of bits corrupted in the two steps together is limited to be at most $pn$. In each step, the decisions made by the channel are based on the information transmitted in and before the step at hand. The notion of (average error) capacity is defined as done above. As explained in the introduction, any lower bound on the capacity of the two-step model holds also for the online channel model.

## 3  Proof of Theorem 1.2

Before presenting the proof of our lower bound for the online channel model, let us start with a short comparison to the Gilbert-Varshamov lower bound that holds for the classical adversarial model. One way to prove the Gilbert-Varshamov bound is to show that a code $C : [2^{Rn}] \to \{0,1\}^n$ chosen at random combined with the nearest neighbor decoder implies a coding scheme of rate almost $1 - H(2p)$ with high probability. Roughly speaking, the achievable rate in this argument is affected by the number of codewords $x$ that are *far away* from any other codeword in $C$. Namely, one is interested in proving that there are lots of codewords $x$, for which the ball of radius $2pn$ centered at $x$ includes no codewords except $x$. Indeed, such a transmitted codeword $x$ will be decoded correctly by a nearest neighbor decoder no matter which error is imposed by the channel. As the volume of this ball is $\sum_{i=0}^{2pn} \binom{n}{i} \sim 2^{H(2p)n}$ the rate essentially follows.

Recall that for our lower bound on the capacity of the online channel we consider the two-step model. In the first step Alice sends a prefix $m$ of length $\alpha n$ and the channel chooses which bits to flip out of these $\alpha n$ bits, and in the second step Alice sends the remaining $(1-\alpha)n$ bits and the channel again chooses which bits to flip out of the remaining part of the codeword. Let us now study the required "forbidden ball" corresponding to a codeword $x$ in the two-step model. To take advantage of the two-step model, consider fixing an error pattern $e$ imposed on the first portion of $x$. Let $\mathcal{B}(x, e)$ be the subset of $\{0,1\}^n$ that satisfies the following property: if the codeword $x$ was transmitted, the error pattern $e$ was imposed on the first portion of $x$ in the first step, and there are no codewords other than $x$ in $\mathcal{B}(x, e)$; then no matter what the channel does in the second step the decoding of Bob will succeed. We define $\mathcal{B}(x, e)$ (denoted as $\mathcal{B}(z)$ for $z = x + e$) rigorously and analyze its size in the upcoming section. Specifically, we show that the size of $\mathcal{B}(z)$ is exponentially smaller than $2^{H(2p)n}$. This fact is a core ingredient in our proof. Combining it with several additional ideas leads to our improved lower bound.

We now turn to present the proof of Theorem 1.2. In Section 3.1 we formally define the "forbidden ball" $\mathcal{B}(z)$ described above and analyze its size. In Sections 3.2 and 3.3 we prove our theorem by showing that with high probability over the codeword $x$ chosen by Alice, the decoding is successful. Namely, that Bob decodes a codeword $x'$ which is equal to the transmitted codeword $x$. In Section 3.2 we analyze the probability (over $x$) that Bob decodes an incorrect codeword $x'$ in which $x$ and $x'$ differ in their first $\alpha n$ bits. In Section 3.3 we address $x$ and $x'$ which agree on their first $\alpha n$ bits. Finally, in Section 3.4 we prove Theorem 1.2.

## 3.1 The "Forbidden Ball" $\mathcal{B}_\alpha^{(p,q)}(z)$

Consider a situation in which Alice transmits a codeword $x$. Namely, in the first step, Alice sends the first $\alpha n$ bits of $x$ and the channel flips $qn$ of them for some $q \in [0, \min(p, \alpha)]$. Let $e_1 \in \{0,1\}^{\alpha n} \times \{0\}^{(1-\alpha)n}$ be the vector of Hamming weight $qn$ that represents the channel's corruptions in the first step, and let $z = x + e_1$ be the (partially) corrupted codeword after the first step. In the second step Alice sends the remaining $(1-\alpha)n$ bits of $x$. Since the channel is limited to a total number of $pn$ corruptions, at most $(p-q)n$ of the bits can be flipped in this step. Let $e_2 \in \{0\}^{\alpha n} \times \{0,1\}^{(1-\alpha)n}$ be the vector of Hamming weight at most $(p-q)n$ that represents the channel's corruptions in the second step, and let $w = z + e_2 = x + e_1 + e_2$ be the corrupted codeword received by Bob.

Conditioning on the first step, namely on the value of $z$, we are interested in counting the vectors that the channel (in its second step) may enforce Bob to consider in his nearest neighbor decoding. These are all the vectors $y \in \{0,1\}^n$ for which there exists a vector $w \in \{0,1\}^n$ such that

- $w$ is of distance at most $pn$ from $y$, and
- $w$ and $z$ agree on the first $\alpha n$ bits and the distance between them is at most $(p-q)n$.

Notice that the second item follows from the fact that our channel can only corrupt bits in the $(1-\alpha)n$ suffix of $z$ in the second step. We define

$$\mathcal{B}_\alpha^{(p,q)}(z) = \{y \in \{0,1\}^n \mid \exists w \in \{0,1\}^n \text{ s.t. } \operatorname{dist}_H(w,y) \leq pn,\ z|_{[\alpha n]} = w|_{[\alpha n]},\ \operatorname{dist}_H(z,w) \leq (p-q)n\}.$$

It is not hard to verify that (a) the original transmitted codeword is in $\mathcal{B}_\alpha^{(p,q)}(z)$, and (b) if this is the only codeword in $\mathcal{B}_\alpha^{(p,q)}(z)$ then Bob will decode successfully. It is also not hard to verify that the size of $\mathcal{B}_\alpha^{(p,q)}(z)$ does not depend on $z$ and therefore we can denote $B_\alpha^{(p,q)} = |\mathcal{B}_\alpha^{(p,q)}(z)|$ for any $z \in \{0,1\}^n$. The following claim bounds $B_\alpha^{(p,q)}$ and is proven in the appendix.

**Claim 3.1.** *For any* $0 < p < \frac{1}{2} \cdot H^{-1}(\frac{1}{2})$ *there exists an* $\eta > 0$ *such that for any* $1 - H(2p) \leq \alpha \leq 1 - 2p$ *and* $q \in [0, p]$ *it holds that* $B_\alpha^{(p,q)} \leq 2^{(H(2p)-\eta)n}$.

## 3.2 Errors Caused by Codewords with Distinct Prefixes

Let $C : [2^{Rn}] \to \{0,1\}^n$ be a code chosen at random and let $x \in \{0,1\}^n$ be a codeword sent by Alice. Consider the setting in which Alice, in the first step, sends the prefix $m = x|_{[\alpha n]}$ and the channel corrupts $qn$ of its bits for some $q \in [0, \min(p, \alpha)]$. Let $e \in \{0,1\}^{\alpha n} \times \{0\}^{(1-\alpha)n}$ be the vector of Hamming weight $qn$ that represents the channel's corruptions in the first step. In the second step Alice sends the last $(1-\alpha)n$ bits of $x$ and the channel is allowed to flip at most $(p-q)n$ of these bits. After the first step, the set of vectors that are of Hamming distance at most $pn$ from a vector that the channel can cause Bob to receive is exactly $\mathcal{B}_\alpha^{(p,q)}(x+e)$. Therefore, if a nearest neighbor decoder fails then there must be another codeword of $C$ (in addition to $x$) in $\mathcal{B}_\alpha^{(p,q)}(x+e)$. In this section we study the probability that $\mathcal{B}_\alpha^{(p,q)}(x+e)$ contains a codeword with a prefix that differs from $m$ and show that it is small no matter what $m$ or $e$ are. Here, the probability is taken over the random construction of $C$.

In general, it is not hard to verify that *in expectation*, indeed a random code $C$ will ensure an exponentially decaying decoding error in the case under study (here, the expectation is over the code construction and the error is over the messages of Alice). However, as the events corresponding to correct decoding are not independent of each other, our proof includes a rather delicate analysis. Our proof in this section consists of two parts. In the first part, we identify a certain property on codes $C$, and prove that it holds with very high probability. This property is then used in the second part of our proof, and enables to cope with the dependencies mentioned above. We start by defining our needed property on $C$.

A code is considered as *good with respect to the pair* $(m, e)$ if it has the following two properties: (a) the number of codewords with prefix $m$ is close to its expectation and, in addition, (b) the number of codewords that do not start with $m$ but alternatively may cause a decoding error on the transmission of a word that does start with $m$ is not much larger than the expectation. This notion is formally defined below. We then show that for every $m$ and $e$ a code $C$ chosen at random is good with respect to $(m, e)$ with high probability. Recall the definitions of $\mathbf{C}^m$ and $\overline{\mathbf{C}^m}$ from Section 2.2.

**Definition 3.2.** *For a natural number $n$, $p > 0$, $R > 0$, $\varepsilon > 0$, $\alpha = R - \varepsilon$, $m \in \{0,1\}^{\alpha n}$ and $e \in \{0,1\}^{\alpha n} \times \{0\}^{(1-\alpha)n}$ of Hamming weight $qn$ for $q \in [0, \min(p, \alpha)]$, we say that a code $\mathbf{C} : [2^{Rn}] \to \{0,1\}^n$ is* good with respect to *the pair $(m, e)$ if*

1. $2^{\varepsilon n - 1} \leq |\mathbf{C}^m| \leq 2^{\varepsilon n + 1}$, *and*

2. $\sum_{x \in Z_m} |\{u \in \overline{\mathbf{C}^m} \mid \mathbf{C}(u) \in \mathcal{B}_\alpha^{(p,q)}(x + e)\}| \leq B_\alpha^{(p,q)} \cdot 2^{\varepsilon n + 2}$,

*where $Z_m$ is the set of all vectors in $\{0,1\}^n$ with $m$ as a prefix, i.e., $Z_m = \{z \in \{0,1\}^n \mid z|_{[\alpha n]} = m\}$.*

A remark regarding Item (2) of Definition 3.2 is in place. In general, Item (2) estimates the number of codewords in $\overline{\mathbf{C}^m}$ that happen to be included in "forbidden balls" of type $\mathcal{B}_\alpha^{(p,q)}(x + e)$ for vectors $x \in Z_m$ (namely, $x|_{[\alpha n]} = m$). Later in our proof, we will think of $x$ as a randomly chosen codeword with prefix $m$, and the l.h.s. of Item (2) will correspond to the expected number of codewords in its "forbidden ball".

**Lemma 3.3.** *For every large enough $n$, $p > 0$, $R > 0$, $\varepsilon > 0$, $\alpha = R - \varepsilon$, a prefix $m \in \{0,1\}^{\alpha n}$ and $e \in \{0,1\}^{\alpha n} \times \{0\}^{(1-\alpha)n}$ of Hamming weight at most $pn$, the probability that a code $C : [2^{Rn}] \to \{0,1\}^n$ chosen at random is good with respect to $(m, e)$ is at least $1 - e^{-2^{\Omega(n)}}$.*

**Proof:** Fix a pair $(m, e)$ and assume that the Hamming weight of $e$ is $qn$ for $q \in [0, p]$. For every $u \in [2^{Rn}]$ denote by $X_u$ the indicator random variable defined to be 1 if $u \in C^m$ and 0 otherwise. Notice that the $X_u$'s are independent and identically distributed and that $|C^m| = \sum_{u \in [2^{Rn}]} X_u$. Also, $\mathbf{E}[X_u] = \Pr[X_u = 1] = \frac{1}{2^{\alpha n}}$, and linearity of expectation implies that $\mathbf{E}[|C^m|] = 2^{Rn} \cdot \frac{1}{2^{\alpha n}} = 2^{\varepsilon n}$. Applying the standard Chernoff bound (see, e.g., [1] Appendix A) we get that Item (1) of Definition 3.2 holds with probability at least $1 - e^{-2^{\Omega(n)}}$.

Now, given that (1) holds, we will show that the probability that (2) holds is $1 - e^{-2^{\Omega(n)}}$. This will imply that with such probability both (1) and (2) hold, as follows from $\Pr[(1) \wedge (2)] = \Pr[(1)] \cdot \Pr[(2)|(1)]$.

Since the summands in Item (2) of Definition 3.2 are not independent, we cannot directly apply the Chernoff-Hoeffding bound. To overcome this issue, we express the summation in (2) as another summation of independent random variables. Details follow. Recall that $Z_m$ stands for the set of all vectors in $\{0,1\}^n$ with $m$ as a prefix. Define for every $u \in \overline{C^m}$ the random variable

$$Y_u = \left|\{x \in Z_m \mid C(u) \in \mathcal{B}_\alpha^{(p,q)}(x + e)\}\right|.$$

Namely, $Y_u$ counts the number of balls $\mathcal{B}_\alpha^{(p,q)}(x + e)$ (with $x \in Z_m$) which include $C(u)$. Denote $Y = \sum_{u \in \overline{C^m}} Y_u$ and observe that $Y$ equals the sum from (2). Observe that the $Y_u$'s are independent and, moreover, they are independent even when conditioning on the size of the set $C^m$. Given that $u \in \overline{C^m}$, for every $x \in Z_m$ the probability that $C(u) \in \mathcal{B}_\alpha^{(p,q)}(x + e)$ is at most $\frac{B_\alpha^{(p,q)}}{2^n - 2^{(1-\alpha)n}}$. Hence,

$$\mathbf{E}[Y_u] \leq |Z_m| \cdot \frac{B_\alpha^{(p,q)}}{2^n - 2^{(1-\alpha)n}} \leq 2 \cdot 2^{(1-\alpha)n} \cdot \frac{B_\alpha^{(p,q)}}{2^n} = \frac{B_\alpha^{(p,q)}}{2^{\alpha n - 1}}.$$

Notice that for every $u \in \overline{C^m}$ we have $Y_u \leq B_\alpha^{(p,q)}$ and define $Y'_u = \frac{Y_u}{B_\alpha^{(p,q)}} \in [0,1]$ and $Y' = \frac{Y}{B_\alpha^{(p,q)}}$. For any $k \in [2^{\varepsilon n-1}, 2^{\varepsilon n+1}]$ use the Chernoff-Hoeffding bound (Theorem 2.3) to obtain

$$\Pr\left[Y \geq B_\alpha^{(p,q)} \cdot 2^{\varepsilon n+2} \,\Big|\, |C^m| = k\right] \leq \Pr\left[Y \geq \frac{4B_\alpha^{(p,q)}(2^{Rn}-k)}{2^{\alpha n}} \,\Big|\, |C^m| = k\right]$$

$$= \Pr\left[Y' \geq \frac{4(2^{Rn}-k)}{2^{\alpha n}} \,\Big|\, |C^m| = k\right] = e^{-\Omega(2^{\varepsilon n})}.$$

Finally, for a large enough $n$ we obtain

$$\Pr[(2)|(1)] = 1 - \sum_{k \in [2^{\varepsilon n-1}, 2^{\varepsilon n+1}]} \Pr\left[Y \geq B_\alpha^{(p,q)} \cdot 2^{\varepsilon n+2} \,\Big|\, |C^m| = k \wedge (1)\right] \cdot \Pr\left[|C^m| = k \,\Big|\, (1)\right]$$

$$\geq 1 - e^{-\Omega(2^{\varepsilon n})} \cdot \sum_{k \in [2^{\varepsilon n-1}, 2^{\varepsilon n+1}]} \Pr\left[|C^m| = k \,\Big|\, (1)\right] = 1 - e^{-\Omega(2^{\varepsilon n})}.$$

∎

We now turn to the second part of our proof. Let $m$ be a prefix of a codeword sent by Alice and let $e$ be the vector that represents the corruptions made by the channel in the first step. Consider a fixed choice of the codewords in $C$ which do not have $m$ as a prefix (i.e., $C|_{\overline{C^m}}$). The following lemma shows that the number of messages in $C^m$ for which the channel may cause a decoding error due to messages in $\overline{C^m}$ is small with high probability. The probability here is over the choice of the codewords that start with $m$ (since $C|_{\overline{C^m}}$ is fixed).

For any $u \in C^m$ define $T_u$ to be the number of codewords of messages from $\overline{C^m}$ in the "forbidden ball" corresponding to $u$. Namely, $T_u = |\{u' \in \overline{C^m} \mid C(u') \in \mathcal{B}_\alpha^{(p,q)}(C(u)+e)\}|$. Let $P_u$ be an indicator random variable defined to be 1 if $T_u \geq 1$ and 0 otherwise. Finally, we let $P^{(m,e)}$ denote the number of codewords with prefix $m$ whose corresponding "forbidden balls" contain codewords associated with elements from $\overline{C^m}$. Formally, $P^{(m,e)} = \sum_{u \in C^m} P_u$. We stress that messages $u$ with $P_u = 1$ are considered as messages for which the channel may cause a decoding error. Thus our objective is to show that $P^{(m,e)}$ is small.

**Lemma 3.4.** *For every $0 < p < \frac{1}{2} \cdot H^{-1}(\frac{1}{2})$ there exists a $\delta_p > 0$ such that for $\varepsilon \leq \delta \leq \delta_p$, $R = 1 - H(2p) + \delta$ and $\alpha = R - \varepsilon$ the following holds for any sufficiently large $n$. For every prefix $m \in \{0,1\}^{\alpha n}$, $e \in \{0,1\}^{\alpha n} \times \{0\}^{(1-\alpha)n}$ of Hamming weight at most $pn$, a fixed set of messages $\overline{C^m}$ and a fixed restriction $\widetilde{\mathbf{C}}$ of $C$ to $\overline{C^m}$,*

$$\Pr\left[P^{(m,e)} < 2^{\varepsilon n/2} \,\Big|\, C|_{\overline{C^m}} = \widetilde{\mathbf{C}} \wedge C \text{ is good with respect to } (m,e)\right] \geq 1 - e^{-2^{\Omega(n)}}.$$

*Here, the probability is taken over the random construction of $C$.*

**Proof:** For $0 < p < \frac{1}{2} \cdot H^{-1}(\frac{1}{2})$ take $\delta_p = \min\left(\frac{4}{7} \cdot \eta, H(2p) - 2p\right)$, where $\eta$ is the constant whose existence is guaranteed in Claim 3.1. Notice that $\delta_p > 0$ since $H(2p) > 2p$, as follows from Fact 2.2.

Fix a pair $(m,e)$ and assume that the Hamming weight of $e$ is $qn$ for $q \in [0,p]$. Denote by $G^{(m,e)}$ the event that $C$ is good with respect to $(m,e)$. Conditioning on $C|_{\overline{C^m}} = \widetilde{\mathbf{C}}$ and on $G^{(m,e)}$, every $C(u)$ for $u \in C^m$ is independently and uniformly distributed over the vectors in $\{0,1\}^n$ that start with $m$, and in particular the $P_u$'s are independent. Since $C$ satisfies Item (2) of Definition 3.2 we get that for every $u \in C^m$,

$$\mathbf{E}\left[P_u \,\Big|\, C|_{\overline{C^m}} = \widetilde{\mathbf{C}} \wedge G^{(m,e)}\right] \leq \mathbf{E}\left[T_u \,\Big|\, C|_{\overline{C^m}} = \widetilde{\mathbf{C}} \wedge G^{(m,e)}\right] \leq \frac{B_\alpha^{(p,q)} \cdot 2^{\varepsilon n+2}}{2^{(1-\alpha)n}} = \frac{4B_\alpha^{(p,q)}}{2^{(1-R)n}}.$$

Notice that our choice of $\delta_p$ implies that $1 - H(2p) \leq R - \varepsilon = \alpha \leq R \leq 1 - 2p$ and hence $B_\alpha^{(p,q)} \leq 2^{(H(2p)-\eta)n}$ by Claim 3.1. Since $C$ satisfies Item (1) of Definition 3.2 we obtain that

$$\mathbf{E}\left[P^{(m,e)} \,\Big|\, C|_{\overline{C^m}} = \widetilde{\mathbf{C}} \wedge G^{(m,e)}\right] \leq |C^m| \cdot \frac{4B_\alpha^{(p,q)}}{2^{(1-R)n}} \leq \frac{8 \cdot 2^{(\varepsilon+H(2p)-\eta)n}}{2^{(1-R)n}}$$

$$= 8 \cdot 2^{(\delta+\varepsilon-\eta)n} \leq 8 \cdot 2^{(\varepsilon/4 + \frac{7}{4}\delta_p - \eta)n} \leq 8 \cdot 2^{\varepsilon n/4}.$$

For a sufficiently large $n$, applying the Chernoff-Hoeffding bound (Theorem 2.3) yields

$$\Pr\left[P^{(m,e)} \geq 2^{\varepsilon n/2} \mid C|_{\overline{C^m}} = \widetilde{\mathbf{C}} \wedge G^{(m,e)}\right] \leq \Pr\left[P^{(m,e)} \geq 16 \cdot 2^{\varepsilon n/4} \mid C|_{\overline{C^m}} = \widetilde{\mathbf{C}} \wedge G^{(m,e)}\right] \leq e^{-2^{\Omega(n)}},$$

as desired. ∎

Combining Lemmas 3.3 and 3.4 we get the following corollary.

**Corollary 3.5.** *For every $0 < p < \frac{1}{2} \cdot H^{-1}(\frac{1}{2})$ there exists a $\delta_p > 0$ such that for $\varepsilon \leq \delta \leq \delta_p$, $R = 1 - H(2p) + \delta$ and $\alpha = R - \varepsilon$ the following holds for any sufficiently large $n$. The probability that a code $C : [2^{Rn}] \to \{0,1\}^n$ chosen at random satisfies that for every prefix $m \in \{0,1\}^{\alpha n}$ and $e \in \{0,1\}^{\alpha n} \times \{0\}^{(1-\alpha)n}$ of Hamming weight at most $pn$, $C$ is good with respect to $(m,e)$ and $P^{(m,e)} < 2^{\varepsilon n/2}$, is at least $1 - e^{-2^{\Omega(n)}}$.*

**Proof:** Let $(m,e)$ be a fixed pair and denote by $G^{(m,e)}$ the event that $C$ is good with respect to $(m,e)$. In the following $\widetilde{\mathbf{C}}$ denotes a restriction of $C$ to $\overline{C^m}$. We have

$$\begin{aligned}
\Pr\left[P^{(m,e)} < 2^{\varepsilon n/2} \wedge G^{(m,e)}\right] &= \sum_{\widetilde{\mathbf{C}}} \Pr\left[P^{(m,e)} < 2^{\varepsilon n/2} \mid C|_{\overline{C^m}} = \widetilde{\mathbf{C}} \wedge G^{(m,e)}\right] \cdot \Pr\left[C|_{\overline{C^m}} = \widetilde{\mathbf{C}} \wedge G^{(m,e)}\right] \\
&\geq (1 - e^{-2^{\Omega(n)}}) \cdot \sum_{\widetilde{\mathbf{C}}} \Pr\left[C|_{\overline{C^m}} = \widetilde{\mathbf{C}} \wedge G^{(m,e)}\right] \\
&= (1 - e^{-2^{\Omega(n)}}) \cdot \Pr\left[G^{(m,e)}\right] \geq (1 - e^{-2^{\Omega(n)}}) \cdot (1 - e^{-2^{\Omega(n)}}) \geq 1 - e^{-2^{\Omega(n)}},
\end{aligned}$$

where the first and the second inequalities follow, respectively, from Lemmas 3.4 and 3.3. Taking the union bound over all the possible pairs $(m,e)$ completes the proof. ∎

### 3.3 Errors Caused by Codewords with the Same Prefix

In this section we consider decoding errors caused by codewords in $C$ that have prefix (of length $\alpha n$) identical to the prefix of the transmitted codeword. Namely, we consider the scenario that Alice sends a codeword $x$, Bob gets the corrupted vector $y$, and the message that Bob outputs corresponds to a codeword that differs from $x$ but shares the prefix $x|_{[\alpha n]}$. A way to handle such errors is to verify that for every prefix $m$, our code $C$ does not include (many) pairs of codewords that share $m$ as a prefix and are *close* together, namely of Hamming distance at most $2pn$. This is the type of analysis that actually corresponds to the classical adversarial channel, and can be used here as we are considering a special case of decoding errors.

The following lemma says that a code $C : [2^{Rn}] \to \{0,1\}^n$ chosen at random with $R < 1 - 4p$ has only few pairs of codewords that share a prefix and have Hamming distance at most $2pn$.

**Lemma 3.6.** *For every $p \in [0, \frac{1}{4})$, $R < 1 - 4p$, a sufficiently small $\varepsilon > 0$ and $\alpha = R - \varepsilon$ there exists a $\gamma > 0$ for which the following holds for any sufficiently large $n$. The probability that a code $C : [2^{Rn}] \to \{0,1\}^n$ chosen at random satisfies that*

1. *for every $m \in \{0,1\}^{\alpha n}$, $2^{\varepsilon n - 1} \leq |C^m| \leq 2^{\varepsilon n + 1}$,*

2. *and for every $m \in \{0,1\}^{\alpha n}$, besides at most $2^{(\alpha - \gamma)n}$ of them, there exists a set $X_m \subseteq C^m$ of size $|X_m| < 2^{(\varepsilon - \gamma)n}$ such that every distinct $u_1, u_2 \in C^m \setminus X_m$ satisfy $\mathrm{dist}_H(C(u_1), C(u_2)) > 2pn$,*

*is at least $1 - e^{-2^{\Omega(n)}}$.*

In order to prove Lemma 3.6 we need the following (known) claim that shows that with high probability a random code almost achieves the Gilbert-Varshamov bound. We include its proof for completeness.

**Claim 3.7.** *For any $p' \in (0, \frac{1}{4})$, $\varepsilon' > 0$ and $R' \leq 1 - H(2p') - \varepsilon'$ the following holds for any sufficiently large $n'$. The probability that for a code $C : [2^{R'n'}] \to \{0,1\}^{n'}$ chosen at random there exists a set $X \subseteq [2^{R'n'}]$ of size $|X| < 2^{(R'-\varepsilon'/2)n'}$ such that every distinct $u_1, u_2 \in [2^{R'n'}] \setminus X$ satisfy $\mathrm{dist}_H(C(u_1), C(u_2)) > 2p'n'$ is at least $1 - 2^{-\frac{3}{8}\varepsilon' n'}$.*

**Proof:** The probability of two distinct messages to be mapped by $C$ to codewords of Hamming distance at most $2p'n'$ is $\frac{\sum_{i=0}^{2p'n'} \binom{n'}{i}}{2^{n'}} \sim \frac{2^{H(2p')n'}}{2^{n'}}$. Denote by $Y$ the number of pairs of messages which are mapped by $C$ to codewords of Hamming distance at most $2p'n'$ and notice that $\mathbf{E}[Y] \lesssim \frac{2^{2R'n'} \cdot 2^{H(2p')n'}}{2^{n'}}$. Apply Markov's inequality to get that

$$\Pr\left[Y \geq 2^{(R'-\varepsilon'/2)n'}\right] \leq \frac{\mathbf{E}[Y]}{2^{(R'-\varepsilon'/2)n'}} \lesssim 2^{-\varepsilon'n'/2}.$$

This implies that with probability at most $2^{-\frac{3}{8}\varepsilon'n'}$ we have $Y \geq 2^{(R'-\varepsilon'/2)n'}$. Taking one message from every pair counted in $Y$, we get the required set $X$. ∎

We now turn to prove our lemma.

**Proof of Lemma 3.6:** The probability that $C$ satisfies Item (1) is at least $1 - e^{-2^{\Omega(n)}}$ as follows from the argument presented in the proof of Lemma 3.3 and a union bound argument over all the possible $m$'s. Now we will show, given (1) holds, that the probability that (2) holds is $1 - e^{-2^{\Omega(n)}}$. This will imply that both (1) and (2) hold with probability $1 - e^{-2^{\Omega(n)}}$.

In order to analyze the probability of (2), let us first fix the size of the image of $C$ for every prefix: for every $m \in \{0,1\}^{\alpha n}$ denote $k_m = |C^m| \in [2^{\varepsilon n - 1}, 2^{\varepsilon n + 1}]$. Denote by $T_m$ the indicator random variable defined to be 1 if there is no set $X_m \subseteq C^m$ of size $|X_m| < 2^{(\varepsilon - \gamma)n}$ such that every distinct $u_1, u_2 \in C^m \setminus X_m$ satisfy $\mathrm{dist}_H(C(u_1), C(u_2)) > 2pn$ (where $\gamma$ is some positive constant to be determined later). In addition, define $T = \sum_{m \in \{0,1\}^{\alpha n}} T_m$. Notice that given the fixed $k_m$'s, we can think of $C$ as $2^{\alpha n}$ random mappings, where the mapping which corresponds to $m$ maps every element in a domain of size $k_m$ to an element in $\{0,1\}^{(1-\alpha)n}$ uniformly and independently. Denote $n' = (1-\alpha)n$, $R' = \frac{\varepsilon}{1-\alpha} \approx \frac{1}{n'} \cdot \log k_m$ and $p' = \frac{p}{1-\alpha}$. Our assumption that $R < 1 - 4p$ implies that $H(2p')$ is bounded away from 1 and hence for a small enough $\varepsilon > 0$ we have that $R' = \frac{\varepsilon}{1-\alpha} \leq 1 - H(2p') - \varepsilon'$ for some $\varepsilon' > 0$. Define $\gamma = \varepsilon'(1-\alpha)/4$. Apply Claim 3.7 and derive that the probability that there is no set $X_m \subseteq C^m$ of size $|X_m| < 2^{(R'-\varepsilon'/2)n'} = 2^{(\varepsilon - 2\gamma)n} < 2^{(\varepsilon-\gamma)n}$ such that every distinct $u_1, u_2 \in C^m \setminus X_m$ satisfy $\mathrm{dist}_H(C(u_1), C(u_2)) > 2pn$ is at most $2^{-\frac{3}{2}\gamma n}$. Therefore, $\mathbf{E}[T_m] = \Pr[T_m = 1] \leq 2^{-\frac{3}{2}\gamma n}$. The $T_m$'s are independent (given the fixed $k_m$'s) so for a sufficiently large $n$ we can apply the Chernoff-Hoeffding bound (Theorem 2.3) to get

$$\Pr\left[T \geq 2^{(\alpha-\gamma)n} \,\Big|\, \forall m.\, |C^m| = k_m\right] \leq \Pr\left[T \geq 2 \cdot 2^{(\alpha - \frac{3}{2}\gamma)n} \,\Big|\, \forall m.\, |C^m| = k_m\right] \leq e^{-2^{\Omega(n)}}.$$

Finally,

$$\begin{aligned}
\Pr[(2)|(1)] &= 1 - \sum_{\{k_m\}_{m \in \{0,1\}^{\alpha n}}} \Pr\left[T \geq 2^{(\alpha-\gamma)n} \,\Big|\, \forall m.\, |C^m| = k_m \wedge (1)\right] \cdot \Pr\left[\forall m.\, |C^m| = k_m \,\Big|\, (1)\right] \\
&= 1 - e^{-2^{\Omega(n)}} \cdot \sum_{\{k_m\}_{m \in \{0,1\}^{\alpha n}}} \Pr\left[\forall m.\, |C^m| = k_m \,\Big|\, (1)\right] = 1 - e^{-2^{\Omega(n)}}.
\end{aligned}$$

∎

## 3.4 Proof of Theorem 1.2

The following corollary stems from Corollary 3.5 and Lemma 3.6 by Fact 2.2 and the union bound.

**Corollary 3.8.** *For every $0 < p < \frac{1}{2} \cdot H^{-1}(\frac{1}{2})$ there exist $\delta > 0$, $\varepsilon > 0$, and $\gamma > 0$ such that for $R = 1 - H(2p) + \delta$ and $\alpha = R - \varepsilon$ the following holds for any sufficiently large $n$. The probability that a code $C : [2^{Rn}] \to \{0,1\}^n$ chosen at random satisfies that*

1. *for every $m \in \{0,1\}^{\alpha n}$, $2^{\varepsilon n - 1} \leq |C^m| \leq 2^{\varepsilon n + 1}$,*

2. *for every prefix $m \in \{0,1\}^{\alpha n}$ and $e \in \{0,1\}^{\alpha n} \times \{0\}^{(1-\alpha)n}$ of Hamming weight at most $pn$, $P^{(m,e)} < 2^{\varepsilon n/2}$,*

3. *and for every $m \in \{0,1\}^{\alpha n}$, besides at most $2^{(\alpha - \gamma)n}$ of them, there exists a set $X_m \subseteq C^m$ of size $|X_m| < 2^{(\varepsilon - \gamma)n}$ such that every distinct $u_1, u_2 \in C^m \setminus X_m$ satisfy $\mathrm{dist}_H(C(u_1), C(u_2)) > 2pn$,*

*is at least $1 - e^{-2^{\Omega(n)}}$.*

Equipped with Corollary 3.8, we are ready to prove Theorem 1.2.

**Proof of Theorem 1.2:** Fix $0 < p < \frac{1}{2} \cdot H^{-1}(\frac{1}{2})$ and let $\delta > 0$, $\varepsilon > 0$, $\gamma > 0$, $R = 1 - H(2p) + \delta$ and $\alpha = R - \varepsilon$ be as in Corollary 3.8. Also, let $\mathbf{C} : [2^{Rn}] \to \{0,1\}^n$ be a code that satisfies the three items in the corollary. Denote by $M$ the set of all $m \in \{0,1\}^{\alpha n}$ for which there is a set $X_m \subseteq \mathbf{C}^m$ of size $|X_m| < 2^{(\varepsilon - \gamma)n}$ such that every distinct $u_1, u_2 \in \mathbf{C}^m \setminus X_m$ satisfy $\mathrm{dist}_H(\mathbf{C}(u_1), \mathbf{C}(u_2)) > 2pn$, and by $\overline{M}$ its complement $\overline{M} = \{0,1\}^{\alpha n} \setminus M$. The corollary guarantees that $|\overline{M}| \leq 2^{(\alpha - \gamma)n}$. We restrict the code $\mathbf{C}$ to the domain $U = [2^{Rn}] \setminus (\cup_{m \in M} X_m)$ and denote the restricted code by $\mathbf{C}' : U \to \{0,1\}^n$. Notice that $|U| \geq 2^{Rn} - 2^{\alpha n} \cdot 2^{(\varepsilon - \gamma)n} = 2^{Rn} - 2^{(R-\gamma)n} \geq 2^{Rn-1}$ for a sufficiently large $n$. We show that this code and the nearest neighbor decoder supply high probability of correct decoding and hence imply the theorem.

Let $x \in \{0,1\}^n$ be the codeword sent by Alice and denote by $m_x = x|_{[\alpha n]} \in \{0,1\}^{\alpha n}$ the vector that Alice sends in the first step of the two-step model. We first show that the probability over Alice's messages that $m_x \in \overline{M}$ is exponentially decaying: $\Pr[m_x \in \overline{M}] = \sum_{m \in \overline{M}} \frac{|\mathbf{C}^m|}{|U|} \leq |\overline{M}| \cdot \frac{2^{\varepsilon n + 1}}{2^{Rn-1}} \leq 2^{(\alpha-\gamma)n} \cdot \frac{2^{\varepsilon n + 1}}{2^{Rn-1}} \leq 2^{-\gamma n + 2}$. Thus, we may neglect the event that $m_x \in \overline{M}$.

Now assume that $m_x \in M$. Observe that for every $m \in M$ the number of codewords of $\mathbf{C}'$ that start with $m$ satisfies $|\mathbf{C}^m \setminus X_m| \geq 2^{\varepsilon n - 1} - 2^{(\varepsilon - \gamma)n} \geq 2^{\varepsilon n - 2}$ for a large enough $n$. In the first step of our two-step model the channel outputs $m_x + e'$ for some $e' \in \{0,1\}^{\alpha n}$ of Hamming weight at most $pn$. Extend $e'$ to a vector $e \in \{0,1\}^n$ by concatenating it to $(1-\alpha)n$ zeros. We now bound the probability of incorrect decoding averaged over all codewords $x$ with prefix $m_x$. We divide our analysis according to the cases discussed in Sections 3.2 and 3.3.

For the analysis corresponding to Section 3.2 consider the probability (taken over messages in $\mathbf{C}^{m_x} \setminus X_{m_x}$) that the "forbidden ball" corresponding to $x$ and $e$ contains a codeword with a prefix that differs from $m_x$. Recall that this probability bounds the probability of a decoding error in the setting of Section 3.2, and, by our definitions, is at most $\frac{P^{(m_x,e)}}{|\mathbf{C}^{m_x} \setminus X_{m_x}|} \leq \frac{2^{\varepsilon n/2}}{2^{\varepsilon n - 2}} \leq 2^{2 - \varepsilon n/2}$. Here, the bound on $P^{(m_x,e)}$ holds since the code satisfies Item (2) in Corollary 3.8.

For the analysis corresponding to Section 3.3, due to our restriction $\mathbf{C}'$ of $\mathbf{C}$ to $U$ and the assumption $m_x \in M$, $x$ is the only codeword with prefix $m_x$ and Hamming distance at most $2pn$ from $x$. Hence, the "forbidden ball" corresponding to $x$ does not contain a codeword with a prefix that equals $m_x$, implying no decoding error in the setting examined in Section 3.3.

Therefore, the probability (taken uniformly over Alice's message $u \in U$) of an incorrect decoding is at most $\Pr[m_x \in \overline{M}] + \Pr[m_x \in M] \cdot 2^{2 - \varepsilon n/2} \leq 2^{-\gamma n + 2} + 1 \cdot 2^{2 - \varepsilon n/2} = 2^{-\Omega(n)}$. All in all, we obtain that the probability of a correct decoding is arbitrarily close to 1 for a sufficiently large $n$, which concludes our proof.

∎

# References

[1] N. Alon and J. H. Spencer. *The probabilistic method*. Wiley-Interscience Series in Discrete Mathematics and Optimization. Wiley-Interscience [John Wiley & Sons], New York, second edition, 2000.

[2] D. Blackwell, L. Breiman, and A. J. Thomasian. The capacities of certain channel classes under random coding. *The Annals of Mathematical Statistics*, 31(3):558–567, 1960.

[3] T. M. Cover and J. A. Thomas. *Elements of information theory, 2nd edition*. Wiley-Interscience, New York, NY, USA, 2006.

[4] I. Csiszár and J. Korner. *Information Theory: Coding Theorems for Discrete Memoryless Systems, 2nd edition*. Akademiai Kiado, New York, NY, 1997.

[5] I. Csiszár and P. Narayan. The capacity of the arbitrarily varying channel revisited: Positivity, constraints. *IEEE Transactions on Information Theory*, 34(2):181193, 1988.

[6] B. K. Dey, S. Jaggi, and M. Langberg. Codes against online adversaries. *In proceedings of the Forty-Seventh Annual Allerton Conference on Communication, Control, and Computing*, 2009.

[7] B. K. Dey, S. Jaggi, M. Langberg, and A. Sarwate. Coding against delayed adversaries. *To appear in proceedings of IEEE International Symposium on Information Theory*, 2010.

[8] E. N. Gilbert. A comparison of signalling alphabets. *Bell Systems Technical Journal*, 31:504–522, 1952.

[9] V. Guruswami and A. Smith. Codes for computationally simple channels: Explicit constructions with optimal rate. *CoRR*, abs/1004.4017, 2010. To appear in FOCS 2010.

[10] W. Hoeffding. Probability inequalities for sums of bounded random variables. *Journal of the American Statistical Association*, 58(301):13–30, March 1963.

[11] S. Jaggi, M. Langberg, T. Ho, and M. Effros. Correction of Adversarial Errors in Networks. In *proceedings of IEEE International Symposium on Information Theory*, pages 1455–1459, 2005.

[12] M. Langberg, S. Jaggi, and B. K. Dey. Binary causal-adversary channels. In *proceedings of IEEE International Symposium on Information Theory*, pages 2723–2727, Piscataway, NJ, USA, 2009. IEEE Press.

[13] A. Lapidoth and P. Narayan. Reliable communication under channel uncertainty. *IEEE Transactions on Information Theory*, 44(6):2148–2177, 1998.

[14] C. McDiarmid. Concentration. In *Probabilistic methods for algorithmic discrete mathematics*, volume 16 of *Algorithms Combin.*, pages 195–248. Springer, Berlin, 1998.

[15] R. J. McEliece, E. R. Rodemich, H. Rumsey, Jr., and L. R. Welch. New upper bounds on the rate of a code via the Delsarte-MacWilliams inequalities. *IEEE Trans. Information Theory*, IT-23(2):157–166, 1977.

[16] L. Nutman and M. Langberg. Adversarial Models and Resilient Schemes for Network Coding. *In proceedings of IEEE International Symposium on Information Theory*, pages 171–175, 2008.

[17] A. Sahai and S. Mitter. The necessity and sufficiency of anytime capacity for stabilization of a linear system over a noisy communication link, Part I: scalar systems. *IEEE Transactions on Information Theory*, 52(8):3369–3395, 2006.

[18] A. Sarwate. Robust and adaptive communication under uncertain interference. *PhD thesis, Berkeley*, 2008.

[19] C. E. Shannon. A mathematical theory of communication. *The Bell System Technical Journal*, 27:379–423,623–656, July, October 1948.

[20] R. R. Varshamov. Estimate of the number of signals in error correcting codes. *Dokl. Acad. Nauk*, 117:739–741, 1957.

# Appendix

## A  Proof of Claim 3.1

First, by our setting of $p$, notice that $\alpha \geq \frac{1}{2} + \gamma$ for some $\gamma \in (0, \frac{1}{2})$ that depends only on $p$. Observe that

$$B_\alpha^{(p,q)} \leq \sum_{k=0}^{(2p-q)n} \sum_{i \leq \min(pn,k)} B_{k,i},$$

where $B_{k,i} = \binom{\alpha n}{i} \cdot \binom{(1-\alpha)n}{k-i}$. Here, with respect to the notation in the definition of $\mathcal{B}_\alpha^{(p,q)}(z)$, we set $k$ to be equal to the Hamming distance between $z$ and $y$, and $i$ to be equal to the Hamming distance between $z|_{[\alpha n]}$ and $y|_{[\alpha n]}$. Notice that $k \leq \mathrm{dist}_H(z,w) + \mathrm{dist}_H(w,y) \leq (2p-q)n$ and as $z|_{[\alpha n]} = w|_{[\alpha n]}$ it holds that $i \leq pn$. In order to prove the claim it suffices to prove the existence of an $\eta > 0$ for which for every $0 \leq k \leq (2p-q)n$ and $i \leq pn$, we have $B_{k,i} \lesssim 2^{(H(2p)-\eta)n}$. Fix such a $k$ and denote $k = (2p-q')n$ for some $q' \in [q, 2p]$. Consider the following two cases:

- Case 1: $q' \geq 2\gamma \cdot p$. Denoting $i = \beta k$ for $\beta \in [0,1]$ we obtain

$$B_{k,i} = \binom{\alpha n}{\beta k} \cdot \binom{(1-\alpha)n}{(1-\beta)k} \sim 2^{\alpha n H(\frac{\beta k}{\alpha n})} \cdot 2^{(1-\alpha)n H(\frac{(1-\beta)k}{(1-\alpha)n})} \leq 2^{nH(\frac{k}{n})} = 2^{nH(2p-q')} \leq 2^{n(H(2p)-\eta_1)},$$

where the first inequality follows from Fact 2.1, and the second holds for $\eta_1 = H(2p) - H((2-2\gamma)p)$ since $H$ is monotonically increasing in $[0, \frac{1}{2}]$ and $2p < \frac{1}{2}$. Notice that $\eta_1$ depends solely on $p$.

- Case 2: $q' < 2\gamma \cdot p$. In this case we have $\alpha(2p - q') > p \cdot 2\alpha(1-\gamma) \geq p(1+2\gamma)(1-\gamma) > p$. Denoting $i = (p-\beta)n$ for $\beta \in [0,p]$ such that $p - q' + \beta \leq 1 - \alpha$ we obtain

$$B_{k,i} = \binom{\alpha n}{(p-\beta)n} \cdot \binom{(1-\alpha)n}{(p-q'+\beta)n} \sim 2^{\alpha n \cdot H(\frac{p-\beta}{\alpha}) + (1-\alpha)n \cdot H(\frac{p-q'+\beta}{1-\alpha})} \leq 2^{\alpha n \cdot H(\frac{p}{\alpha}) + (1-\alpha)n \cdot H(\frac{p-q'}{1-\alpha})}.$$

To verify the last inequality, one can show that $\alpha(2p-q') > p$ implies that the function $g : [0,p] \to [0,1]$ defined by $g(\beta) = \alpha \cdot H\left(\frac{p-\beta}{\alpha}\right) + (1-\alpha) \cdot H\left(\frac{p-q'+\beta}{1-\alpha}\right)$ is monotonically decreasing, as follows from calculating its derivative. The assumption $\alpha \leq 1 - 2p$ implies that $H\left(\frac{p-q'}{1-\alpha}\right) \leq H\left(\frac{p}{1-\alpha}\right)$ since $H$ is monotonically increasing in $[0, \frac{1}{2}]$. Now, let $\alpha^*$ be the $\alpha \in [\frac{1}{2} + \gamma, 1 - 2p]$ that maximizes $\alpha \cdot H(\frac{p}{\alpha}) + (1-\alpha) \cdot H(\frac{p}{1-\alpha})$ [2] and obtain that

$$B_{k,i} \lesssim 2^{\alpha^* n \cdot H(\frac{p}{\alpha^*}) + (1-\alpha^*)n \cdot H(\frac{p}{1-\alpha^*})} = 2^{n(H(2p)-\eta_2)},$$

where the equality holds for some $\eta_2 > 0$ that depends solely on $p$, as follows from Fact 2.1 using $\frac{p}{\alpha^*} < \frac{p}{1-\alpha^*}$.

Choosing $\eta = \min(\eta_1, \eta_2)$ completes the proof.

---

[2] It can be seen that $\alpha^* = 1 - 2p$.